\documentstyle[preprint,aps,eqsecnum,floats,epsfig]{revtex}

\newcommand{\beq}{\begin{equation}}
\newcommand{\eeq}{\end{equation}}
\newcommand{\beqa}{\begin{eqnarray}}
\newcommand{\eeqa}{\end{eqnarray}}

\newcommand{\la}{\mbox{$\lambda$}}
\newcommand{\lp}{\mbox{$\lambda^\prime$}}

\newcommand{\gsim}{\mbox{${~\raise.15em\hbox{$>$}\kern-.85em
	  \lower.35em\hbox{$\sim$}~}$}}
\newcommand{\lsim}{\mbox{${~\raise.15em\hbox{$<$}\kern-.85em
	  \lower.35em\hbox{$\sim$}~}$}}

\def\nnu{\nonumber}

\def\sh{\hat{s}}

\def\ea{{\it et al.\/}}

\def\npb#1{Nucl.\ Phys.\ {\bf B #1}}
\def\plb#1{Phys.\ Lett.\ {\bf B #1}}
\def\prd#1{Phys.\ Rev.\ {\bf D #1}}
\def\prl#1{Phys.\ Rev.\ Lett. {\bf #1}}

\begin{document}

\draft

{\tighten

\preprint{\vbox{\hbox{UdeA-PE-98/001} 
                \hbox{DIFUB-98-11} 
		\hbox{hep-ph/9806359} }}

\title{ $B \to \tau\, \mu\> (X)$ decays in SUSY models 
without R-parity}

\author{Dafne Guetta\,$^{a,1}$, Jesus M. Mira\,$^{b,2}$ 
and Enrico Nardi\,$^{b,3}$ }

\footnotetext{
\begin{tabular}{ll}
& $^1$E-mail address: guetta@bo.infn.it \\
& $^2$E-mail address:  jmira@pegasus.udea.edu.co \\ 
& $^3$E-mail address: enardi@fisica.udea.edu.co 
\end{tabular} }

\address{ \vbox{\vskip 0.truecm}
  $^a$Dipartimento di Fisica -- 
Universit\'a degli Studi di Bologna \\
Via Irnerio 46, I-{\it 40126}  \  Bologna, \ Italy \\
\vbox{\vskip 0.truecm}
  $^b$ Departamento de F\'\i sica -- 
Universidad de Antioquia \\
A.A. {\it 1226}, \ Medell\'\i n,  \ Colombia }

\maketitle

\begin{abstract}%
Being strictly forbidden in the standard model, experimental detection
of the lepton flavor violating decays $B\,(\bar B) \to \tau^+\, 
\mu^-\,$ and $b\,(\bar b) \to X\, \tau^+\, \mu^-\,$ would constitute
an unmistakable indication of new physics. We 
study these decays in supersymmetric models 
without R-parity and  without lepton number. In 
order to derive order of magnitude predictions for 
the branching ratios, we assume a horizontal 
$U(1)$ symmetry with horizontal charges chosen to 
explain the magnitude of fermion masses and quark 
mixing angles. We find that the branching ratios 
for decays with a $\tau\mu$ pair in the final state 
are not particularly suppressed 
with respect to the lepton flavor 
conserving channels. In general in these models 
 ${\rm B}[b \to 
\mu^+\,  \mu^-\,  (X)] \lsim 
{\rm B}[b (\bar b) \to \tau^+\, \mu^-\,  (X)] \lsim 
{\rm B}[b \to \tau^+\, \tau^-\, (X)]\,$.  
While in some cases the rates for final states 
$\tau^+ \tau^-\,$ can be up to one order of 
magnitude larger than the lepton flavor violating 
channel, due to  better efficiencies for muon 
detection and to the absence of standard model contributions, 
decays into   $\tau\mu$ final 
states appear to be better suited to reveal 
this kind of new physics. 
\end{abstract}
} 
\bigskip
\leftline{ PACS number(s):  13.20.-v, 13.20.He, 12.60.Jv, 11.30.Hv}

{\tighten   

\newpage
\section{Introduction}

In the standard model (SM) the $SU(2)\times U(1)$ 
gauge symmetry together with Lorentz invariance 
implies  accidental Baryon (B) and Lepton (L) 
number conservation at the renormalizable level. 
Due to the larger Lorentz 
structure, supersymmetric (SUSY)  versions of the 
SM allow for renormalizable B and L violating 
operators involving scalars with non-zero B and 
L charges, that can induce fast proton decay as 
well as several other unobserved processes. 
Therefore, additional symmetries are required to 
enforce proton stability and to suppress B and 
L violating transitions. In most SUSY models, 
invariance under the additional  parity quantum 
number ${\rm R}=(-1)^{3{\rm B} + {\rm L} + 2S}$ 
($S$ being the 
spin) is assumed, and this enforces B and L 
conservation at the renormalizable level. However, 
today it is believed that B and L are not 
likely to be fundamental symmetries of nature, and 
in fact a much larger spectrum of models it is 
known to be consistent  with the data. To render 
phenomenologically viable SUSY extensions of the 
SM, the first priority is to ensure proton 
stability. In this respect other symmetries can 
be more effective than R parity, since R parity still  
allows for potentially dangerous dimension five 
B and L violating 
operators. Some interesting alternatives exist 
which forbid dimension four and five B violating 
terms, and hence are more effective to ensure 
proton stability \cite{Ross}. Since in these 
models L number can be violated by 
renormalizable operators, they imply a quite different 
phenomenology from R-parity conserving SUSY 
models \cite{susynoR}. 

Two new types of Lagrangian  terms characterize 
this class of models.  First we have bilinear 
terms which couple the three lepton doublets superfields  
$\hat L_e\,,\hat L_\mu\,$ and $\hat L_\tau$ with  the up-type 
Higgs superfield $\hat \Phi_u$. The main effect of these 
terms is to induce neutrino masses via 
neutrino-neutralino mixing 
\cite{HallSuzuki,BGNN1,Rnumass,RGnu}. Requiring 
that the resulting masses do not exceed the 
experimental limits, implies constraints on the 
structure of these models \cite{HallSuzuki,BGNN1,Rnumass,RGnu}.
In the models we will study these constraints are automatically 
satisfied, thanks to the presence of a horizontal symmetry 
that suppresses all the contributions to neutrino masses. 
Another effect of the bilinear terms is that of mixing fermions in
different representations of $SU(2)$. In turn, this can generate 
flavor changing couplings of the $Z$ boson to the leptons. 
In our theoretical framework also these  effects are  safely 
suppressed below the experimental sensitivity. 
For these reasons the effects of the bilinear terms do no warrant
further elaboration in the present context. 

Secondly we have a set of renormalizable  interactions 
in the superpotential which are responsible for 
L and lepton flavor (${\rm L}_i\,$, $i=e\,,\mu\,,\tau$) violating transitions. 
In the mass basis, these terms read 
\beq \label{Lviolmass}
\la_{i j k} \> \hat L_i\, \hat L_j\,  \hat {l^c_k} +
\lp_{i j k} \> \hat L_i\,\hat Q_j\,\hat {d^c_k}\,, 
\eeq
where $\hat Q_i $ and $ \hat { d^c_i} $ denote the quark doublet and
down-quark singlet superfields, $\hat { l^c_i}$ are the 
lepton  singlets, 
and  $\la_{i j k}=-\la_{j i k} $ due to the antisymmetry 
in the $SU(2)$ indices.

Several of the $\la$ and $\lp$ couplings are 
strongly constrained by the existing phenomenology 
\cite{susynoR}. The best limits are for couplings 
involving fermions of the first two generations 
($i,j,k=1,2$) while for couplings involving more 
than a single third generation field the existing 
limits are much weaker. From the theoretical point 
of view, the values of the $\la$ and $\lp$ 
couplings in (\ref{Lviolmass}) are not predicted 
by the model. However, general models that can 
explain the observed fermion mass hierarchy also 
predict that R-parity violating couplings 
involving more than a single third generation 
field are the largest ones. This suggests 
that this kind of new physics can be 
effectively searched for in rare decays involving 
$b\,$,  $\tau$ and $\nu_\tau$ \cite{btautau,bsnn,btau}. 
In the following, we will make this statement more 
precise, by imposing on the models additional 
theoretical constraints. Following 
Ref.~\cite{btau} we embed into the R-parity violating 
model a particular horizontal symmetry  that can 
account for the order of magnitude of the fermion 
masses and CKM  angles 
\cite{horizontal,YossiYuval}. This framework allows us 
to estimate the size of the  
relevant L violating couplings in 
(\ref{Lviolmass}). 

A rather complete study of $B$ decays into third 
generation leptons like $B\to \tau\bar\nu_\tau\,,$ 
$B\to \tau^+ \tau^-\,,$ $b\to X\tau^+ \tau^-\,,$ 
$b\to X \nu_{\tau} \bar\nu_{\tau}$  has been recently presented 
in Ref.\cite{btau}. The sensitivity of these decay 
modes to new physics from SUSY models without 
R-parity was thoroughly investigated, and 
compared to the sensitivity of the corresponding 
decay modes into $\mu,\,\nu_{\mu}.$ It was found that the 
processes that are most sensitive to these effects 
are the leptonic decays $B_{d,s} \to \tau^+\tau^-$ 
and the inclusive decay  $b \to X_s 
\tau^+\tau^-\,$. However, from the experimental 
point of view the efficiency for $\tau$ 
identification is expected to be rather low at 
$B$-factories. Similarly, 
$\tau$-tagging will be a very hard task at 
future high $B$ statistics
experiments at hadron colliders \cite{experimental}.  
  This constitutes a serious drawback for  
the search of new physics effects in decays with 
final state $\tau$'s, and it is unlikely that 
the theoretical enhancement of the decay rates
could fully compensate for this. 

In this paper we  point out that the lepton flavor 
violating decays $B_{d,s} (\bar B_{d,s}) 
\to \tau^+\mu^-$  and $b\, (\bar b)\to X_s 
\tau^+\mu^-\,$ together with the corresponding 
$CP$ conjugate decays, can provide the best 
compromise between the two requirements of large 
theoretical branching ratios and  good 
efficiencies in searching for the experimental 
signatures. Indeed, for the two body decay $B \to 
\tau^+\mu^-$ the absence of a $\mu^+$ with 
momentum opposite to the $\mu^-$ in the $B$ rest 
frame represents a clean signature, rather easy 
to search for. The first experimental limit on this 
decay  ${\rm B}(B\to \tau^\pm\mu^\mp) < 8.3\times 10^{-4}$ 
has been recently established by the CLEO 
Collaboration \cite{CLEOBtaumu}.
The search for the three body decay $b \to X_s 
\tau^+\mu^-\,$ appears to be more difficult,  
because of the lack of knowledge of the 
momentum of the missing  $\mu^+.$  
This is reflected in the present experimental situation. 
While a tight limit on the lepton flavor violating 
decay $b \to X_s \mu^+e^-\,$ has been recently established
[${\rm B}(b \to X_s \mu^\pm e^\mp)\,< 2.2\times 10^{-5}$
\cite{CLEObXmumu}]
to date no experimental limit exists on  
decays into $X_s \tau^+ e^-$ or $X_s \tau^+ \mu^-$ final states. 

At hadron colliders, already the study of the two body decay 
 $B\rightarrow \tau^{+} \mu^{-}$ will be  
 a difficult task. This is because in this case 
the momentum  of the decaying $B$ is not known, 
hence the presence of large  backgrounds,  
as for example from the  semileptonic 
decays $B\rightarrow D^{(*)} \mu   \bar \nu \,$, 
will render quite challenging the search for this rare decay. 

From the theoretical point of view, 
the detection of lepton flavor violating decays
would represent a striking evidence of  physics
beyond the SM. The absence of SM contributions, and 
in the case of the decay 
$b \to X_s \tau^+\mu^-\,$ 
the absence of long distance effects 
which are difficult to  estimate in a reliable way \cite{LDnew},  
render these decays  well 
suited to reveal in a clean way new physics effects.   
In summary, because of the large theoretical  
 enhancement of the branching ratios with respect 
 to $ \mu^{+}\mu^{-}$ final states, and since in 
 any case a muon is experimentally much easier to identify than a  
 tau, we believe that these processes will allow 
to search for signals of SUSY models without R parity 
with a better sensitivity 
than  lepton  flavor conserving decays
into $ \mu^{+}\mu^{-}$ or $ \tau^{+}\tau^{-}$. 

In the next section we will outline the main 
features of SUSY models without R-parity 
embedded in models with horizontal symmetries. In 
Sec.~III we will present the relevant 
expressions for the effective new physics 
coefficients which appear in these models, and for 
the decay rates. Finally, in Sec.~IV we will 
discuss our results and present our conclusions. 
 
\section{ R-parity violation 
		    in the framework of horizontal symmetries}
\vspace{0.1in}

In order to evaluate  the effects of the 
R-parity violating interactions in (\ref{Lviolmass}), 
we need to estimate quantitatively
the coefficients $\lambda$ and $\lambda^\prime$. 
We work in the framework of the supersymmetric  models 
with horizontal symmetries that have been  
thoroughly investigated in Refs.\cite{horizontal,YossiYuval}.
We assign to each supermultiplet $\hat \psi$  a  charge 
$H(\hat\psi)$ of an Abelian  horizontal group ${\cal{H}}
= U(1)_{H}\,$ which is explicitly broken by a small 
parameter $\varepsilon$ with charge $H(\varepsilon)=-1\,$. This    
gives rise to a set of selection rules for the effective couplings 
appearing in the low  energy  Lagrangian \cite{horizontal}.  
Assuming that each of the lepton, quark and Higgs superfields 
carries a positive or zero  charge, the selection rule  
relevant for the present  discussion is that the effective coupling 
$g_{abc}$ for a general trilinear superpotential term 
$\hat \psi_a\hat \psi_b\hat \psi_c$ is of order 
$g_{abc}\sim \varepsilon^{H(\hat\psi_a)+H(\hat\psi_b)+H(\hat\psi_c)}\,$.
Therefore, the leptons and down-type quarks Yukawa couplings are 
respectively  of order 
$Y^l_{ij}\sim \varepsilon^{H(\hat\Phi_d)+H(\hat L_i)+H(\hat l^c_j)}$ and 
$Y^d_{ij}\sim \varepsilon^{H(\hat\Phi_d)+H(\hat Q_i)+H(\hat 
d^c_j)}\,$. 
Most of the L-violating couplings in (\ref{Lviolmass})  
are further suppressed with respect to the corresponding 
Yukawa couplings. They can be  estimated as
\beq \label{lanum}  
\lambda_{kij} \sim Y^l_{ij} \> \varepsilon^{H(L_k) - H(\Phi_d)}  
	    \sim \left(\frac{2 \sqrt{2} G_F}{\cos^2\beta}\right)^{1/2} \> 
\> m_{l_i}\>{\varepsilon}^{H(l^c_j) - H(l^c_i) + H(L_k) - 
H(\Phi_d)}\,,  
\eeq
and
\beq  \label{lpnum}
\lambda^\prime_{kij} \sim Y^d_{ij} \varepsilon^{H(L_k) - H(\Phi_d)} 
	  \sim \left(\frac{2 \sqrt{2} G_F}{\cos^2\beta}\right)^{1/2} \> 
 m_{d_i}\> {\varepsilon}^{H(d^c_j) - H(d^c_i) + H(L_k) - H(\Phi_d)}\,,
\eeq
where $G_F$ is the Fermi constant, and $\tan\beta= 
\langle\Phi_u\rangle/\langle\Phi_d\rangle $ with $\Phi_u$ the 
up-type Higgs doublet. 
From Eqs.(\ref{lanum}) and (\ref{lpnum}) it is apparent that 
in our framework the 
couplings $\la$ and $\lp$  involving fermions of the  third 
generation are respectively enhanced by $m_\tau$ and $m_b\,$. 

\noindent
In order to give a numerical estimate of the couplings, 
we need a set of $ H$ charges  and a value for  $\varepsilon\,$. 
The magnitude of the ${\cal{H}}$ breaking parameter 
is generally taken to be the value of the Cabibbo angle, 
$\varepsilon\sim 0.22\,,$ while the quark, lepton and 
Higgs  charges are chosen to reproduce the values of the fermion masses 
and CKM mixing angles. Besides reproducing the measured values, 
the model  has some predictivity in the quark sector \cite{horizontal},  
it yields estimates for ratios of neutrino masses \cite{YossiYuval,BGNN2} 
and, most important in the present context, it ensures that the 
L-violating couplings in (\ref{Lviolmass}), (\ref{lanum}) and (\ref{lpnum}) 
are safely suppressed below the present experimental limits \cite{BGNN1}. 
The following $\cal H$-charge assignments  fit the order 
of magnitude of all the quark masses and CKM mixing angles
\cite{horizontal}:  
\beqa\label{Hquarks}
  \hat Q_{1}     \  \ \hat Q_{2}     \   \ \hat Q_{3}           &\qquad&   
     \  \hat d^c_{1} \ \  \hat d^c_{2} \ \ \hat d^c_{3}   \qquad 
     \  \hat u^c_{1} \ \ \hat u^c_{2} \ \ \hat u^c_{3}   \qquad
     \  \hat \Phi_{d}    \ \ \hat \Phi_{u} 
\nnu\\
 (3) \ (2) \ (0) &\qquad& 
 (3) \ (2) \ (2)  \qquad 
 (3) \ (1) \ (0)   \qquad
 \ (0) \ (0) \,.
\eeqa
Following Ref.~\cite{btau}, we use for the leptons two 
different sets of charges which define two different models, 
and for each model we chose  a different value of the squark masses 
$ m_{\tilde q} \,$:
\beqa\label{Hleptons}
 \hat   L_{1}    \   \ \hat L_{2}     \   \ \hat L_{3}        &\qquad& 
 \ \hat l^c_{1}\ \ \, \hat l^c_{2} \ \ \, \hat l^c_{3}   \qquad 
\qquad m_{\tilde l}\>({\rm GeV}) \quad \ m_{\tilde q}\>({\rm GeV})\nnu\\ 
{\rm Model\ I: \ }\qquad\qquad  (4) \ (2) \ (0)    &\qquad&  
  (4) \ (3) \ (3)    \qquad 
\qquad \ 100   \qquad\qquad   170    \nnu\\
{\rm Model\ II:} \qquad\qquad  (3) \ (0) \ (0)    &\qquad&  
  (5) \ (5) \ (3)    \qquad 
\qquad \ 100   \qquad\qquad   350   \,.
\eeqa
Model I tends to enhance new physics 
effects induced by operators 
arising from squark exchange, while in model~II the effects of  
new scalar operators induced by slepton exchange 
tend to dominate. 
The choices (\ref{Hquarks}) and (\ref{Hleptons}) for the horizontal charges 
are not unique. Since the Yukawa interactions are invariant  
under a set of  $U(1)$ symmetries such as B, L and 
hypercharge,
it is  always possible to shift the $H$-charges of any amount 
proportional to one of the corresponding $U(1)$ quantum numbers without 
affecting the predictions for the masses and mixing 
angles. In particular, the shift (proportional to L)   
$H(\hat L_i)\to H(\hat L_i) + n\,$, $H( \hat l^c_i) 
\to H( \hat l^c_i) - n\,$ and 
$H(\hat\psi) \to H(\hat\psi)$ for all the other fields 
has the effect of suppressing (for $n>0\,$) all the L violating couplings 
in (\ref{Lviolmass}), (\ref{lanum}) and (\ref{lpnum})   
by a factor of $\varepsilon^n\,$.   
It turns out that already for $n=1$  the 
suppression is large enough so that all
 the lepton flavor violating 
decays will be unobservable at most of the future $B$-physics 
experiments. 
We also notice that model~II can be derived from model~I by means of  shifts  
proportional to lepton flavor numbers with $n_e = -1\,$, $n_\mu  = -2\,$, 
$n_\tau = 0\,$. This has the effect of enhancing some of the $\la$ 
couplings without affecting the charged lepton masses. 
Of course, the predictions for the neutrino 
mixing angles will be different in the two models.

\section{Coefficients and Rates for the decays }

In the models under investigation, 
the lepton flavor violating decay $\bar B_q \to \tau^+ \mu^- $
\break
(with $q=d,s$) and $b \to X_{s} \tau^+ \mu^-$ 
are induced by the effective Lagrangian 
\beqa \label{LbarBmum} 
- {\cal L}_{\rm eff}^{-}  
 & = & 
C_{1\, S}^{-}\, \left(\bar{q}_{L} b_{R}\right)
\left(\bar{\mu}_{R} \tau_{L}\right) +
 C_{2\, S}^{-}\, \left(\bar{q}_{R} b_{L}\right)
\left(\bar{\mu}_{L} \tau_{R}\right) 
\nnu\\
& + &
C_{V}^{-}\, \left(\bar{q}_{R}\gamma^{\mu} b_{R}\right)
\left(\bar{\mu}_{L} \gamma_{\mu}\tau_{L}\right) + h.c.\,. 
\eeqa
The first two operators arise from 
sneutrino exchange diagrams, while the last one 
corresponds to squark exchange diagrams after 
Fierz rearrangement. 
The coefficients appearing in  (\ref{LbarBmum}) read
\beq\label{Cminus} 
C_{1\, S}^{-} = \sum_{i\neq 3}\, 
\frac{{\lambda^\prime}^*_{iq3}\lambda_{i32}}
	   {  m^{2}_{\tilde \nu_i}}\,; \qquad 
C_{2\, S}^{-}\  =\sum_{i\neq 2}\, 
\frac{\lambda^{\prime}_{i3q}{\lambda}^*_{i23}}
	   {  m^{2}_{\tilde \nu_i}} \,; \qquad  
C_{V}^{-} = \sum_{i} 
\frac{{\lambda^\prime}^*_{2i3}\lambda^{\prime}_{3iq}}
{2\, m^2_{\tilde q_i} }  \,;
\eeq
where the index values $i=3$ in $C_{1\, S}^{-} $ 
and $i=2$ in $C_{2\, S}^{-} $ 
are excluded because of the antisymmetry 
in the first two indices of the $\lambda$ couplings.

For the decays $B_q \to \tau^+ \mu^-$ 
and $\bar b \to X_{s} \tau^+ \mu^-$ 
the effective Lagrangian reads 
\beqa \label{LbarBmup}
- {\cal L}_{\rm eff}^{+} 
& = & 
C_{1\, S}^{+}\, 
\left(\bar{b}_{L} q_{R}\right)
\left(\bar{\mu}_{R} \tau_{L}\right) +
 C_{2\, S}^{+}\, \left(\bar{b}_{R} q_{L}\right)
\left(\bar{\mu}_{L} \tau_{R}\right) 
\nnu\\
& + &
C_{V}^{+}\, \left(\bar{b}_{R}\gamma^{\mu} q_{R}\right)
\left(\bar{\mu}_{L} \gamma_{\mu}\tau_{L}\right) + h.c.\,, 
\eeqa
		     where 
\beq\label{Cplus}
C_{1\, S}^{+} = \sum_{i\neq 3}
\frac{{\lambda^\prime}^*_{i3q}\lambda_{i32}}
	   {  m^{2}_{\tilde \nu_i}} \,; \qquad 
C_{2\, S}^{+} \ =\ 
\sum_{i\neq 2} \frac{\lambda^{\prime}_{iq3}{\lambda}^*_{i23}}
	   {  m^{2}_{\tilde \nu_i}} \,; \qquad 
C_{V}^{+}  = \sum_{i}
\frac{{\lambda^\prime}^*_{2iq}\lambda^{\prime}_{3i3}}
{2\, m^2_{\tilde q_i} }\,.  
\eeq
The expressions for the various branching ratios 
are presented in the next two subsections.
In order to simplify the formulae 
 we have neglected the muon mass 
(however, $m_\mu \neq  0$ has been kept
in the numerical analysis).  

\subsection{The decays $\bar B\to \tau^+\mu^-$ and 
$B\to \tau^+\mu^-$ } 

The amplitude for the decay 
$\bar B_q\rightarrow \tau^+\mu^- $  
can be written as 
\beqa \label{Amplm}
{\cal A}^-_q  &=& i f_{B_q}\, m_B\, \frac{1}{4} \,
 \Bigg\{
  \left[\frac{m_{B}}{m_{b}}\,(C_{2\, S}^{-}-
  C_{1\, S}^{-})-\frac{m_{\tau}}{m_{B}}
  C_{V}^{-}\right] (\bar{\mu}\tau)  
  \nnu \\
  &+&
  \left[\frac{m_{B}}{m_{b}}\,(C_{2\, S}^{-}+
  C_{1\, S}^{-})-\frac{m_{\tau}}{m_{B}}
  C_{V}^{-}\right] (\bar{\mu}\gamma_{5} \tau)
   \Bigg\}, 
\eeqa
where we have used the PCAC (partial conserved axial current) relations
$ \langle 0|\,\bar u\,\gamma^\mu\gamma_5\,b\,|\bar B\rangle 
= i f_B\, p^\mu_B \,$ and 
$ \langle 0|\, \bar u\,\gamma_5\,b\, |\bar B\rangle 
\simeq  - i f_B\, m^2_B/m_b \, $.
This yields the branching ratio 
  \beqa \label{BRminus}
{\rm B}(\bar{B}_{q}\rightarrow \tau^{+}\mu^{-} ) &=&
f_{B_q}^{2} \tau_B \frac{m_{B}^{3}\, }{64\,
\pi}\left( 1-\frac{ m^2_\tau}{m^2_B}\right)^3 
\> \Bigg[ 
\Big|\frac{m_{B}}{m_{b}}C_{2\, S}^{-}-
 \frac{m_\tau}{m_B}C_{V}^{-}\Big|^2+
\frac{m^2_{B}}{m^2_{b}}|C_{1\, S}^{-}|^2
\Bigg].
\eeqa
Eq.~(\ref{BRminus}) also accounts for the 
$CP$ conjugate decay $B_q \rightarrow \tau^-\,\mu^+ \,$,
therefore 
experimental searches for both the decay channels
$\bar{B}\rightarrow \tau^{+}\mu^{-}$ and 
$B \rightarrow \tau^{-}\mu^{+}$
will yield informations on the same set of operators. 
The decay mode $ B\rightarrow \tau^{+}\mu^{-} $
is controlled by the coefficients $C^+_{1S},
\,C^+_{2S}$ and $C^+_{V}$ in 
(\ref{Cplus}). The amplitude is given by 
$ {\cal A}^+_q  = 
- {\cal A}^-_q  (C^+_{1\,S},C^+_{2\,S},C^+_{V})$ 
with ${\cal A}^-_q $ defined as in 
(\ref{Amplm}).  Therefore the branching ratios 
${\rm B}(B\rightarrow \tau^{+}\mu^{-} ) $ and 
$ {\rm B}(\bar{B} \rightarrow \tau^{-}\mu^{+} )\,$ 
are again given by (\ref{BRminus}) with the substitution 
$\{C^-_{1\,S},\,C^-_{2\,S}\,, C^-_{V}\} \to 
\{C^+_{1\,S},\,C^+_{2\,S}\,, C^+_{V}\} $.

\subsection{The decays $b\to X \tau^+\mu^-$ and 
$\bar b \to X \tau^+\mu^-$ } 

For the double differential distribution  for the inclusive decay 
$b(p)\to s(p^\prime )\, \tau^+(k^\prime )\, \mu^-(k) $ 
with respect to the invariants 
$x=p^\prime \cdot k/m_b^2$ and $y=p \cdot k/m_b^2\,$, we find   
\beqa\label{double}
\frac{d^2\Gamma^{-} }{dx\,dy} & = & \frac{m_b^5}{16\pi^3}\Bigg[ 
4|C_V^-|^2\left(\frac{1-\hat{m}^2_\tau-\hat{m}^2_s}{2}-y\right)y 
\nonumber \\  
& \ & -\left(
|C_{1S}^-|^2+|C_{2S}^-|^2\right)\left(x-y + \frac{1 
-\hat{m}^2_\tau +\hat{m}^2_s }{2} \right)(x-y)
-2Re(C_{2S}^-C_V^{-*})\hat{m}_\tau x
\Bigg].  
\eeqa
where $\hat m_f= m_f/m_b\,$, and  $x$ and $y$ 
range between the following limits 
\beq\label{limits} 
0 \le 
 y  \le \frac{1-(\hat{m}_s +\hat{m}_\tau)^2}{2}\,, 
\qquad \qquad 
x_{-}\le  x  \le x_{+}\,, 
\nonumber \eeq
with 
\beq
x_{\pm}  =  \frac{1}
{2\hat{s}}\left[\hat{m}^2_s+ (1-\hat{s} )\Big(\hat{s} 
-\hat{m}^2_\tau \pm \lambda^{1/2}(\hat{s},\hat{m}^2_s,
\hat{m}^2_\tau)\Big)\right], 
\nonumber \eeq 
where
  $\hat{s} = 
(p-p^\prime)^2/m_{b}^2$ and $\lambda(x,y,z)=x^2+y^2+z^2-2xy-2xz-2yz.$ 

The corresponding expressions for the distribution 
$ d^2\Gamma^{+} /dx\,dy$ describing the decay 
$\bar{b}\rightarrow \bar{s} \, \tau^+ \, \mu^- $
can be obtained from (\ref{double}) by interchanging 
$p \leftrightarrow -p'$  
which yields $x\leftrightarrow -y$
and  $\hat{m}^2_s \leftrightarrow 1$ in the  terms 
inside square brackets, and by substituting 
$\{C^-_{1\,S},\,C^-_{2\,S}\,, C^-_{V}\}$ with 
$\{C^+_{1\,S},\,C^+_{2\,S}\,, C^+_{V}\} $.

We now introduce the  forward-backward asymmetries
$A^\pm_{\rm FB}$ of the two distributions 
$\Gamma^{\pm}\,$. The asymmetries are defined  
with respect to the angular variable
$c_\theta=\cos\theta,$ where $\theta$ is the angle 
between the $\mu^{-}$ momentum $\vec{k}$ and the 
$s$ momentum $\vec{p^\prime}$ in the $B$ rest frame:
\beq \label{Afb}
A^\pm_{\rm FB}(y) = \frac{1}{d\Gamma^\pm (y)/dy} 
\left[
\int_0^1\,dc_\theta \frac{d^2\, \Gamma^\pm
(y,c_\theta) }{dy\, d c_\theta} - 
\int_{-1}^0\,dc_\theta \frac{d^2\, \Gamma^\pm
(y,c_\theta)}{dy\, d c_\theta}
\right]\,.
\eeq
As we will see, the dependence of the asymmetries with respect 
to the normalized muon energy $y=E_\mu/m_b$ can provide important
informations on the underlying new physics, which are 
complementary to the measurements of the branching ratios.  
The average values of the asymmetries are computed as 
\beq\label{AvAsymm}
\langle A_{\rm FB}^\pm \rangle = \frac{1}{\Gamma^\pm}\, 
\int\,dy \frac{d\, \Gamma^\pm(y)}
{dy}\,A^\pm_{\rm FB}(y)\,,
\eeq
where $ \Gamma^\pm = \int\,dy \frac{d\, \Gamma^\pm (y)}{dy}\,$. 
Finally, we also study the total $\mu^-$ asymmetry 
$A_{\rm FB}$ of the  distribution 
$d^2\, [\Gamma^- (y,c_\theta) +\Gamma^+ (y,c_\theta)] /
dy\, d c_\theta$ which turns out to be a useful quantity 
when untagged $B$ samples are used for the measurements. 


\section{Discussion } 

In this section we discuss the numerical predictions for the branching
ratios for the decays $B_{d,s}\,(\bar{B}_{d,s}) \to \tau^+ \mu^-,$ 
 $b\,(\bar b) \to X_s \tau^+ \mu^-\,$ and for the $\mu^-$ 
forward-backward asymmetries measurable in the decays 
into three body final states. 
In our estimates, we have used the following 
set of values for the relevant SM parameters: 
$ m_b=4.8\ {\rm GeV},\ m_s=200\ {\rm MeV},\ 
   m_{\mu} = 106\ {\rm MeV},\ m_{\tau} = 1.777\ {\rm GeV}, 
f_{B_d} = 200\, {\rm MeV}, \ f_{B_s} = 230\, {\rm MeV},\ 
    m_B = 5.3\, {\rm GeV},\ \tau_B=1.6\, {\rm ps}\, $ and 
$ {\rm B}(B\to X_c\,l\bar{\nu})= 10.4\%\,$, while  
the magnitude of the various new physics coefficients 
is  determined 
by the sets of $H$-charges and SUSY masses listed in 
 (\ref{Hquarks}) and (\ref{Hleptons}).  

Our results are collected in Table~I and in Figs.~1-3.  
Table~I lists the numerical results 
for $B$ decays involving the channels $b \to \mu^-$ and  
$\bar b \to \mu^-\,$, which are respectively controlled by 
the two different sets of coefficients  
$\{C^-_{1S,2S,V}\}$ and $\{C^+_{1S,2S,V}\}\,$.  
These coefficients are evaluated in the two different models 
defined by the charges in  (\ref{Hquarks}) and (\ref{Hleptons}).  
The entries in the first column in Table~I 
refer to model~I. We recall that in this case 
the choice of the leptonic horizontal charges 
tends to enhance the effects of squark exchange.  
The entries in the second column refer to model~II. Here squark
exchange diagrams are suppressed by a different choice of the
horizontal charges and by the relatively large value of the squark
masses, so the effects of slepton exchange tend to dominate. 
We stress that the aim of the numerical predictions 
given in the first two columns in Table~I is that 
of suggesting the level of precision that future
$B$-physics experiments will have to reach in 
order to detect, or to
effectively constrain, new physics from 
SUSY without R-parity. 
In the last column we list the existing experimental
limits. It is apparent that  most of the decay modes we have studied 
are presently still unconstrained. 

The first
four lines in Table~I collect the results for the two body leptonic decays, 
while the results for the three body final states are given in 
the following two lines. 
Next we present the results for the average values of the $\mu^-$
forward-backward asymmetries $A_{\rm FB}^\pm$  corresponding 
respectively to
$b\to \mu^-$ and $\bar b\to \mu^-$ decays. The average value of the
untagged $\mu^- $ asymmetry $A_{\rm FB}$  is given in the last line.

%
\begin{table}[t]
\begin{center}
 {\tighten
\caption{\baselineskip 16pt 
Predictions for the branching ratios for the
lepton flavor violating $B$ and $\bar B$ decays 
into $\tau^+\mu^-$ and for the forward-backward 
$\mu^-$ asymmetries, in the R-parity
violating models  discussed in the text.
Model~I is defined by 
the lepton horizontal charges $ H(\hat L) = (4 \,, 2 \,, 0) \,$,
$ H(\hat   l^c) = (4 \,, 3 \,, 3) $  and by the 
SUSY masses $m_{\tilde l} =100$~GeV, $m_{\tilde q} =170$~GeV.
Model~II corresponds to the  horizontal charges 
$ H(\hat L)  = (3 \,, 0 \,, 0) \,$, $ H(\hat l^c ) = (5 \,, 5 \,, 3)$
and to $m_{\tilde l} =100$~GeV and  $m_{\tilde q} =350$~GeV.
In both models the value of the horizontal symmetry breaking parameter 
is $\varepsilon = 0.22\,$. The existing experimental limits are 
given in the third column.}
\label{table3}
\vspace{0.3cm}
\begin{tabular} {c  c  c  c }
Process  &  Model 1   &  Model 2 & 90\% c.l. limit [Ref.] \\
\hline 
\noalign {\vspace{1truemm} }   
$ {\rm B}\ (B_s\to \, \tau^+\mu^-)
	     $&$ 8.3\times 10^{-9}$
& $  7.9\times 10^{-7} $ & \\
$  {\rm B}\ (B_d\to \, \tau^+\mu^-)
	     $&$\, 3.0\times 10^{-10}$
& $   2.9\times 10^{-8} $ & $ 8.3\times 10^{-4} \cite{CLEOBtaumu}$ \\
$ {\rm B}\ (\bar{B}_s\to \, \tau^+\mu^-)
	     $&$ 5.0\times 10^{-7}$
& $  2.7\times 10^{-4} $ & \\
$ {\rm B}\ (\bar{B}_d\to \, \tau^+\mu^-)
	     $&$ 1.8\times 10^{-8}$
& $  1.0\times 10^{-5} $ & $ 8.3\times 10^{-4} \cite{CLEOBtaumu} $\\
\noalign {\vspace{1truemm} }
\hline
\noalign {\vspace{1truemm} }
$ {\rm B}\ (b \to X_s\, \tau^+\mu^-)  
	   $&  $ 1.9\times 10^{-7}$
& $  6.4 \times 10^{-5} $ & \\
$ {\rm B}\ (\bar b\to X_s\, \tau^+\mu^-)  
	   $&  $ 1.6 \times 10^{-7}$
& $  4.1 \times 10^{-6} $ & \\
\noalign {\vspace{1truemm} }   
  \hline
\noalign {\vspace{1truemm} }   
$ \langle {A^-_{\rm FB}}\ (b) \rangle  
	   $&  $ -0.02$
& $  -0.24 $ & \\
$ \langle {A^+_{\rm FB}}\ (\bar b) \rangle  
	   $&  $ -0.75$
& $  -0.71 $ & \\
$ \langle {A_{\rm FB}}\ (b, \bar{b}) \rangle  
$&  $ -0.36$
& $  -0.27 $ & \\
\noalign {\vspace{1truemm} }   
\end{tabular}
 }     
\end{center}
\end{table}
%


To put in evidence the advantage of studying the lepton flavor
violating decay modes with respect to  the 
lepton flavor conserving  decays $b\to \tau^+\tau^-
(X)$  and $b\to \mu^+\mu^- (X)\,$, we list in Table II (taken from 
Ref.~\cite{btau}) the numerical predictions 
for these  decays  as 
derived in the SM, in model~I and in model~II.

%
\begin{table}[t]
\begin{center}
 {\tighten
\caption{\baselineskip 16pt 
Predictions taken from Ref.\protect\cite{btau}
for the lepton flavor conserving $B$ decays 
into $\tau^+\tau^-$ and $\mu^+\mu^-$, in the 
standard model and in the presence of new physics.
Model~I and model~II  coincide with the two models 
of Table~I, and are discussed in the text. 
The existing experimental limits are listed in the 
last column.} 
\label{table1}
\vspace{0.3cm}
\begin{tabular} {c  c  c  c  c }
Process & Standard Model  &  Model 1   &  Model 2 & Limit \\
  \hline
\noalign {\vspace{1truemm} }
$ {\rm B}\ (B_s\to \, \tau^+\tau^-)
	     $&$    9.1 \times 10^{-7} $&$ 5.7\times 10^{-6}$
& $  1.8\times 10^{-4} $ & $ 5.0 \times 10^{-2}$ 
\tablenotemark[1] \cite{btautau}  \\
$ {\rm B}\ (B_d\to \, \tau^+\tau^-)
	     $&$    4.3 \times 10^{-8} $&$ 1.9\times 10^{-7}$
& $  6.3\times 10^{-6} $ & $ 1.5 \times 10^{-2} $ 
\tablenotemark[1]  \cite{btautau} \\
\noalign {\vspace{1truemm} }
\hline
\noalign {\vspace{1truemm} }
$ {\rm B}\ (B_s\to \, \mu^+\mu^-)
	 $&$    4.3 \times 10^{-9} $&$ 7.9\times 10^{-7}$
& $  7.2\times 10^{-7} $ & $ 2.6 \times 10^{-6} $
\tablenotemark[2] \cite{CDFmumu}  \\
$ {\rm B}\ (B_d\to \, \mu^+\mu^-)  
$&$\,  2.1 \times 10^{-10} 
$&$ 2.9\times 10^{-8} $&$   2.7\times 10^{-8} $ 
                          & $ 8.6 \times 10^{-7} $ 
\tablenotemark[2] \cite{CDFmumu} \\ 
\noalign {\vspace{1truemm} }
  \hline
\hline
\noalign {\vspace{1truemm} }   
$ {\rm B}\ (b\to X_s\, \tau^+\tau^-)_{\scriptscriptstyle\rm no\> cut}  
	   $&$ 4.9 \times 10^{-6} $&$ 7.3\times 10^{-6}$ 
& $  7.9 \times 10^{-6} $ & \\
$ {\rm B}\ (b\to X_s\, \tau^+\tau^-)_{\scriptscriptstyle\sh>0.6}  
	   $&$ 1.5 \times 10^{-7} $&$ 2.2\times 10^{-6}$
& $   2.7 \times 10^{-6} $ & $ 5.0 \times 10^{-2} $
\tablenotemark[1] \cite{btautau}  \\
\noalign {\vspace{1truemm} }   
\hline 
\noalign {\vspace{1truemm} } 
$ {\rm B}\ (b\to X_s\, \mu^+\mu^-)_{\scriptscriptstyle\rm no\> cut}  
	     $&$ 3.1 \times 10^{-4} $&$ 3.1\times 10^{-4}$
& $  3.4 \times 10^{-4} $ &  \\  
$ {\rm B}\ (b\to X_s\, \mu^+\mu^-)_{\scriptscriptstyle\sh<0.4 }   
	     $&$ 4.3 \times 10^{-6} $&$ 4.4 \times 10^{-6}$
& $   8.4 \times 10^{-6} $ & $ 5.8 \times 10^{-5} $ 
\tablenotemark[3] \cite{CLEObXmumu}  \\
\noalign {\vspace{1truemm} }   
\end{tabular}
\leftline{
\tablenotemark[1]{\ Limit estimated 
from the non-observation of large missing energy events at 
LEP.} } 
\leftline{
\tablenotemark[2]{\ 95\% c.l.}  \quad 
\tablenotemark[3]{\ 90\% c.l.} }
 }     
\end{center}
\end{table}
%

Our results for the two body decays are as follows: the rates for
$\bar B$ decays are more than two orders of
magnitude larger than the corresponding  $B$ decays. This is
mainly due to the fact that the dominant contribution to the 
two body decays of $\bar B$ mesons is given by 
the coefficient $C^-_{2S}\,$,  
while for the $B\,$,  $C^+_{2S}$ is suppressed by 
the $H$-charge difference 
$H(\hat d^c_3)-H(\hat Q_3) = 2,$ so that it gives negligible 
contributions to the decay rates.  
This is interesting, because -- modulo generation independent shifts
proportional to baryon number or hypercharge -- the set of charges in
(\ref{Hquarks}) is unique for fitting the quark masses and mixing
angles. Therefore, the hierarchy in the pattern of 
the decay rates shown in Table~I can be taken as a general qualitative 
prediction of SUSY
models without R-parity embedded in models with an $U(1)$ horizontal
symmetry. Confronting the figures of the leading $\bar B_q\to 
\tau^+\mu^-$ decay rates with the corresponding figures in 
Table~II, we see that while in model~I the rates are much smaller than
the rates for $B_q \to \tau^+\tau^-,\, $ in model~II there is no 
similar suppression. In this case the larger phase space available 
for the lepton flavor violating decays yields 
a slight enhancement of the rates. 
Confronting with the experimental limit on this decay mode
 given in the third column in Table~I,  
we see that 
an improvement of two orders of magnitude is 
needed in order to test the model.  Such an improvement can be within the reach
of forthcoming $B$-factory experiments. 

In contrast to the two body decays, 
the decay rates for $b$ and $\bar b$
into three body final states 
predicted in model~I are comparable in size. 
This is because the leading contribution 
to both decays now comes
from vector operators, and confronting (\ref{Cminus})
with (\ref{Cplus})   it is
easy to check that $C_V^-=C_V^+ $. Model~II tends to enhance the
effect of scalar operators, and this again
results in a relative enhancement of the $ b$ with respect to the
$\bar b\,$ decay rates. 
A confrontation with the figures given in Table~II 
shows that
in model~II the lepton flavor violating $b$ decay can be 
up to one order of magnitude larger than the rate for 
$b \to X_s \tau^- \tau^+$.  
Again, this is mainly due to phase space effects.  
Table~II also shows that without kinematic cuts, 
the SM rates for three body final states are comparable with the rates
predicted by the new physics models. Then, in 
order to single out   the
new physics short distance effects
in the lepton flavor conserving channels, 
it is necessary to impose suitable cuts.
Table~II shows the effects of two different cuts 
on the $\tau^+\tau^-$ and $\mu^+\mu^-$ invariant
masses.  
Clearly, one of the advantages of studying 
lepton flavor violating channels is the complete absence  
of this background to the new physics effects. 

In spite of some large enhancements from the new physics, 
the figures in Table~I make apparent 
that even in the most favorable cases,  
the predicted branching ratios remain rather small. 
Therefore, in order to measure the corresponding rates, or to put 
significant constraints on the models,
a large statistics and a good experimental efficiency are required. 
It is worth noticing  that while a separate {\it measurement} of the 
relevant combinations of the new physics 
coefficients $\{C^-_{1S,2S,V}\}$ and $\{C^+_{1S,2S,V}\}$ 
requires  the identification of the flavor of the 
decaying  $B\,$, 
with a corresponding loss in the experimental efficiency,   
in order to establish {\it limits}  on 
the new physics coefficients it is sufficient 
to search for $\tau\mu$  pairs production in the decays of  
untagged $B$ samples. This procedure can ensure a 
large gain in statistics and will yield the  strongest bounds. 
If $\tau\mu$ pairs {\it are} detected in the decay of a sample of
untagged $B$'s, then a measurement of the $\mu$ forward-backward asymmetry
could provide the additional information needed to identify the
flavor of the initial state. 

Figs.~1 and 2 depict respectively the forward-backward $\mu^-$
asymmetries for $b$ and  $\bar b$ as a function of the
normalized muon energy in the $B$ rest frame $y=E_\mu/m_b\,$. 
The solid lines correspond 
to model~I, while the dashed lines refer to model~II. 
We see that while the shape of the asymmetries  
is quite similar for the two models, there are large 
differences in the $\mu^-$  angular 
distributions for $b$ or $\bar b$ initial states. 
Namely, the asymmetry for decaying $\bar b$ 
is negative in the whole energy range, yielding the 
large negative averages  listed in Table~I.
In contrast, for decaying $b$'s the asymmetry changes sign.  
This induces large cancellations in the energy averages 
as is apparent from the figures in the next-to-last line in 
Table~I. 
Fig.~3 depicts the asymmetry for an untagged sample of an equal
number of $b$ and $\bar b$ initial states.   
Since in model~II (dashed line) the decay rate for 
$b \to X_s \tau^+ \mu^- $ dominates
over the rate rate for $\bar b\,$, 
the asymmetry for the untagged sample reproduces 
quite closely the energy dependence of the
asymmetry in Fig.~1 (namely there is a change in the  sign). 
In model~I the two decay modes have comparable branching
ratios, and accordingly it is not possible to identify in a reliable way 
the flavor of the initial state just from an inspection 
of the untagged asymmetry (solid line in Fig.~3).  
From these results it is clear that a measurement of a change of sign 
in the energy dependence of the $\mu^-$ asymmetry 
in an untagged $B$ sample would signal that 
most of the $\tau^+ \mu^-$ events  originates 
from decays of $b$'s, while the measurement 
of a negative asymmetry over the whole energy range
would suggest that the contribution from 
decaying $\bar b$'s is at least comparable in size.  

To check to what extent this remarkable feature of
the asymmetries depends on our particular models, 
we have arbitrarily varied the values of the 
scalars and  vector coefficients in the two models. 
Our results indicate that in the limit of 
very heavy squarks $m_{\tilde q} \gg 350$~GeV, 
which implies $C_V^\pm \ll C_S^\pm\,$,   
the difference in the energy dependence 
of the two asymmetries tends to be washed out:  
namely also for initial $\bar b$ we find a sign inversion.  
In the opposite limit $C_S^\pm \ll C_V^\pm\,$,
corresponding to very heavy sleptons,   
both the asymmetries become negative over the whole 
energy range, even if there are large differences 
in the shapes. 
However, if the difference between the 
squarks and sleptons masses is kept within a few 
hundreds GeV, the qualitative features of the 
energy dependence of the asymmetries are maintained, 
rendering possible in principle the identification of the 
flavor of the initial state.

In conclusion, in this paper we have investigated 
$b$ and $\bar b$ lepton flavor violating decays involving 
a $\tau\mu$ pair in the final state. These decays are expected 
to occur in SUSY models without R-parity. 
Assuming a $U(1)$  
horizontal symmetry with fermion charges chosen to explain 
a known set of parameters ( the quark masses, the CKM mixing angles,  
and the lepton masses) allowed us to derive order of 
magnitude predictions for the 
various branching ratios.  
A straightforward prediction 
of this theoretical framework 
is that new physics  effects are stronger 
in decays like   $b\to \tau^+\tau^- (X)$ when 
several third generation fermions are involved.
However, our results indicate that in general 
decays involving a $\tau\mu$ pair in the final state 
are not particularly suppressed 
with respect to the decays involving a pair of $\tau$'s.
On the other hand, due to the presence of a single muon 
in the decay products and to the absence of any SM contribution  
(and in particular of the backgrounds from long distance effects) 
the lepton flavor violating channels are experimentally much easier 
to be searched for. 
Therefore, the decays $\bar B (B) \to \tau^+\mu^-$ and 
$b (\bar b) \to X_s \tau^+\mu^-\,$, together 
with the $CP$ conjugated decays, 
appear to be better suited than the lepton flavor conserving decays 
to search for signals of R-parity violation.

\acknowledgements
We thank Zoltan Ligeti for 
useful comments on the manuscript. 
D. G. acknowledges the Particle Theory Group of the 
University of Antioquia, where this research was carried out,  
for the nice hospitality and for the pleasant working atmosphere.

  } 


\newpage

{\tighten

}  


\newpage

\begin{figure}[t]
\epsfxsize=10 truecm
\epsfysize=7  truecm
\centerline{\epsffile[40 48 263 213]{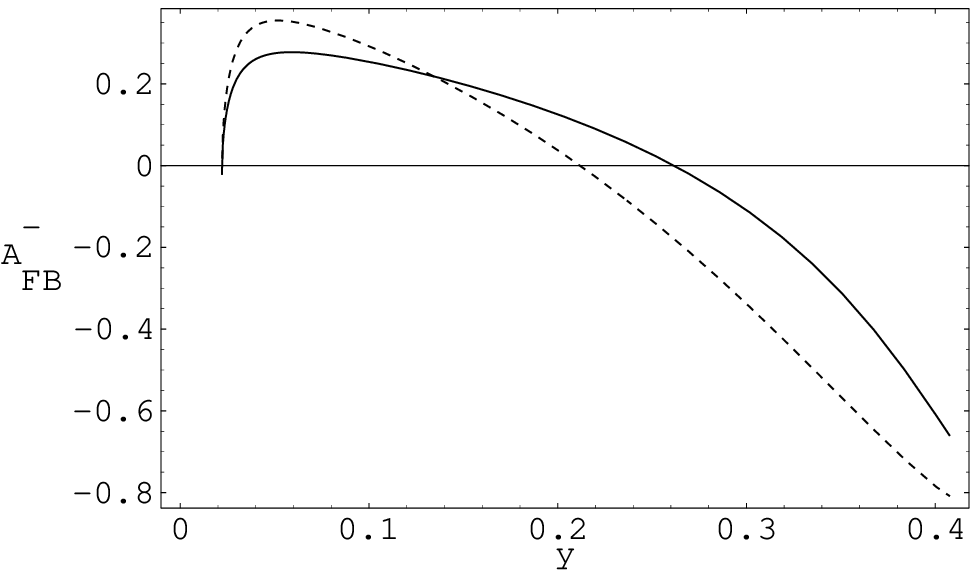}}
 {\tighten
\caption{\baselineskip 16pt 
Predictions for the $\mu^-$ forward-backward asymmetry 
$ {A^-_{\rm FB}}(y) $ in the decay $ b\to X_s\, \tau^+\mu^- $  
as a function of the normalized $B$  rest frame  muon energy 
$y={E_{\mu}}/{m_{b}}$.   
The solid (dashed) line correspond to model~I~(II) 
discussed in the text.
Model~I is defined by 
the lepton horizontal charges $ H(\hat L) = (4 \,, 2 \,, 0) \,$,
$ H(\hat   l^c) = (4 \,, 3 \,, 3) $  and by the 
SUSY masses $m_{\tilde l} =100$~GeV, $m_{\tilde q} =170$~GeV.
Model~II corresponds to the  horizontal charges 
$ H(\hat L)  = (3 \,, 0 \,, 0) \,$, $ H(\hat l^c ) = (5 \,, 5 \,, 3)$
and to $m_{\tilde l} =100$~GeV and  $m_{\tilde q} =350$~GeV.
In both models the value of the horizontal symmetry breaking parameter 
is $\varepsilon = 0.22\,$.} 
 }     
\end{figure}

\begin{figure}[b]
\epsfxsize=10 truecm
\epsfysize=7  truecm
\centerline{\epsffile[40 48 263 213]{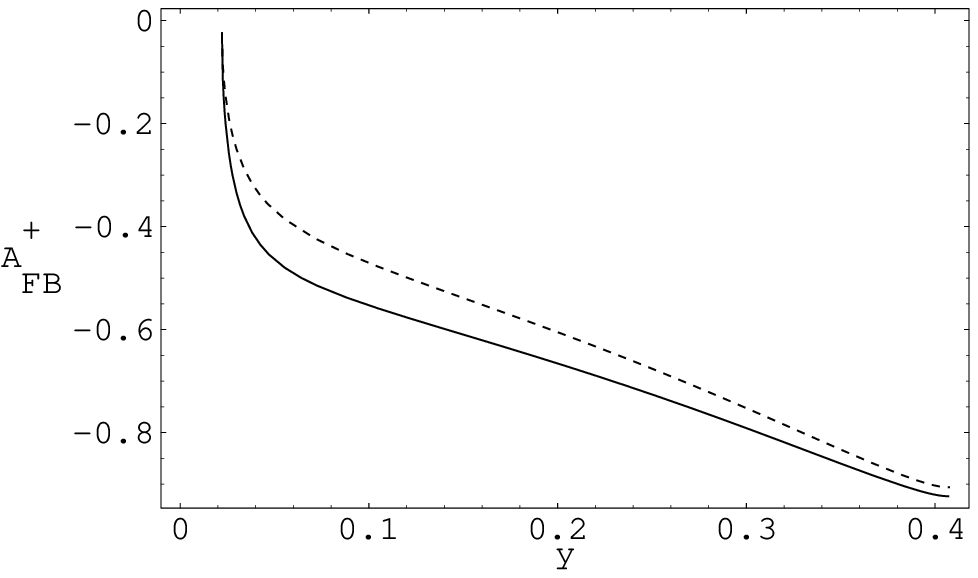}}
 {\tighten
\caption{\baselineskip 16pt 
Predictions for the $\mu^-$ forward-backward asymmetry 
$ {A^+_{\rm FB}}(y) $ in the decay $ \bar b\to X_s\, \tau^+\mu^- $  
as a function of the normalized $B$  rest frame  muon energy 
$y={E_{\mu}}/{m_{b}}$.
The solid (dashed) line correspond to model~I~(II) 
discussed in the text.
The new physics model parameters are as in Fig.~1.}
 }     
\end{figure}

\begin{figure}[t]
\epsfxsize=10 truecm
\epsfysize=7  truecm
\centerline{\epsffile[40 48 263 213]{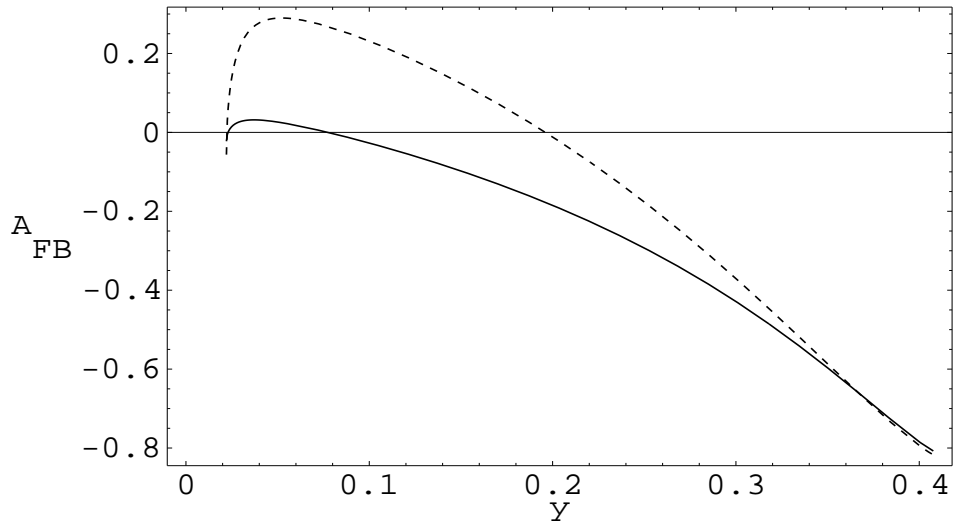}}
 {\tighten
\caption{\baselineskip 16pt 
Predictions for the $\mu^-$ forward-backward asymmetry 
${A_{\rm FB}}(y) $ in the decay of an untagged $b$ 
sample into $X_s\, \tau^+\mu^-$   
as a function of the normalized $B$  rest frame muon energy 
$y={E_{\mu}}/{m_{b}}$. 
The solid (dashed) line correspond to model~I~(II) 
discussed in the text.
The new physics model parameters are as in Fig.~1.}
 }     
\end{figure}

\end{document}